\documentclass[pre,twocolumn,
 amsmath,amssymb,
 aps,
floatfix,
]{revtex4-1}

\usepackage{graphicx}
\usepackage{dcolumn}
\usepackage{bm}
\usepackage{hyperref}

\usepackage{float}
\usepackage{xcolor}
\usepackage{transparent}

\begin{document}

\title{Particle Size Effects in Flow-Stabilized Solids}

\author{Scott Lindauer}
\author{Carlos P. Ortiz}
\author{Robert Riehn}
\author{Karen E. Daniels}
\affiliation{Dept. of Physics, North Carolina State University, Raleigh, NC 27675 USA}

\date{\today}

\begin{abstract}
Flow-stabilized solids are a class of fragile matter that forms when a dense suspension of colloids accumulates against a semi-permeable barrier, for flow rates above a critical value. In order to probe the effect of particle size on the formation of these solids, we perform experiments on micron-sized monodisperse spherical polystyrene spheres in a Hele-Shaw geometry. 
We examine the spatial extent, internal fluctuations, and fluid permeability of the solids deposited against the barrier, and find that these do not scale with the P\'eclet number. Instead, we find distinct behaviors at higher P\'eclet numbers, suggesting a transition from thermal- to athermal-solids which we connect to particle-scale fluctuations in the liquid-like layer at the upstream surface of the solid. We further observe that while the Carman-Kozeny model does not accurately predict the permeability of flow-stabilized solids, we do find a new scaling which predicts the permeability.
\end{abstract}

\maketitle

\section{Introduction}

In many contexts, the separation of colloidal particles into a solid-like phase is either beneficial, for concentrating the material, or detrimental due to creating clogs. 
This has included the glass transition \cite{Gokhale2016}, rheological suspension properties \cite{Hsiao2017,Chen2010}, particle confinement \cite{Teich2016,Zhang2016}, bulk elasticity \cite{Lele2011,Dettmer2014}, and porosity \cite{Molnar2015}. An ability to control this separation is important in industrial settings such as microfiltration for wastewater treatment \cite{Bratby2006,Drogui2016}, detergents \cite{Yonglei2010}, and paints/inks \cite{Park2006}. In many cases, it is the coupling between fluid flow and particle settling which drive the formation of colloidal solids. 

Filter fouling provides a powerful illustration of the diversity of issues involved in this process. It is known that the size of the particles plays a crucial role in microfiltration fouling, as smaller particulate matter more readily diffuses away from membrane surfaces \cite{Bacchin2006}. As such, there have been a number of studies focusing on colloidal fouling \cite{Hoek2003,Cohen1986,Hong1997,Tang2009} and have further prompted detailed studies on how particle size
\cite{Hwang2006,Gassara2008,Tarabara2002,Pavanasam2011,Sripui2011} and particle size distribution \cite{Chang1996,Chang2008,Sripui2011} affects flux loss in filters. Specifically, monodisperse suspensions increase filter flux as particle size increases, while the opposite occurs for polydisperse suspensions \cite{Sripui2011}. Membrane fouling can be characterized by the filtration flux, in which there is net deposition of particles onto the filtration membrane due to advection by the fluid flow occurring more quickly than the diffusion of the particles away from the membrane. The ratio of diffusive to advective timescales is known as the P\'eclet number ($\mathit{Pe}$). The critical flux \cite{Field1995,Howell1995,Wu1999} is defined as the flow velocity below which filtration flux doesn't decline. 
Above the critical flux, the P\'eclet number provides a prediction for circumstances under which microfiltration fouling is expected \cite{Bhattacharya1997,Bacchin2004,Bacchin2006,Zamani2014}. It is therefore important to distinguish which effect arises due to particle size, flow velocity, or the nondimensional P\'eclet number.

\begin{figure}
	\includegraphics[width=\linewidth]{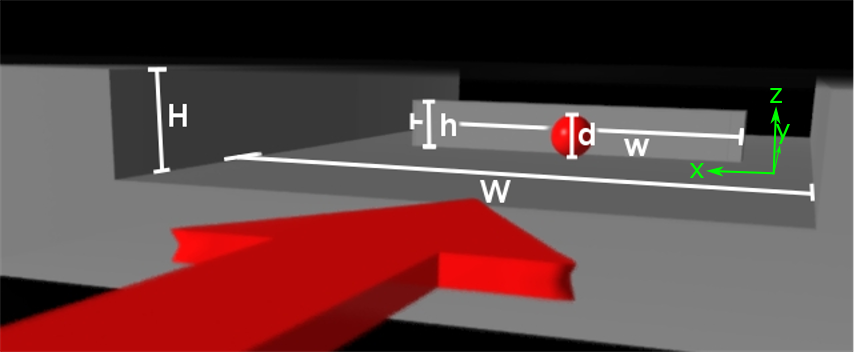}
	\caption[Schematic of microfluidic geometry]{Schematic of microfluidic geometry (not to scale), portraying a single microsphere in front of a barrier perpendicular to fluid flow. The channel geometry is defined by the ratio of channel height $H$ and barrier height $h$, while $W$ is significantly larger than $w$ such that wall effects are ignored.}
	\label{fig:channel}
\end{figure}

In order to examine the relationship between the P\'eclet number and the formation of colloidal solids, we utilize a well-developed experimental system for creating flow-stabilized solids (FSS) \cite{Ortiz2013,Ortiz2014,Ortiz2016}. We create FSS by flowing a dilute colloidal suspension against a semi-permeable barrier within in a microfluidic channel, as illustrated in Fig.~\ref{fig:channel}. The deposition of particles is governed both by the normal and tangential forces arising from the flow field, as well as by thermal diffusion. Normal forces deposit particles directly on the barrier, tangential forces shear particles off of the barrier, and thermal diffusion allows particles to escape the barrier. In the experiments presented here, we utilize polystyrene spheres of diameter $d=0.52\,\mathrm{\mu m}$ to $1.04\,\mathrm{\mu m}$. As in our prior studies \cite{Ortiz2013}, we find that steady-state piles form above a critical P\'eclet number. However, we find that neither the size $A$ of the piles, nor their angle of repose $\theta$ scales universally with $\mathit{Pe}$. 

We isolate these effects as arising from a particle-size dependence which is not completely accounted for by the P\'eclet number. By examining the temporal correlations of particle-fluctuations within the liquid-like layer on the upstream edge of the FSS, we find that two timescales are present: a longer timescale associated with diffusion, and shorter timescale associated with advection. These timescales allow us to classify two types of behavior in the FSS. For P\'eclet number ${\mathcal O}$(1) we observe {\it thermal} FSS, where both timescales are of similar magnitude, and the angle of repose of the pile depends on both P\'eclet number and $d$. For P\'eclet number $>{\mathcal O}$(10) we observe {\it athermal} FSS in which there are two distinct timescales and the angle of repose is approximately independent of $d$.

Finally, we investigate the effectiveness of the Carman-Kozeny (C-K) model for permeability in our system through observations of our system's flow profile. Previous work has identified failures of the C-K model, which include systems of wide void size distribution \cite{Zhang2019} and non-uniform microstructure \cite{Hill2001,Hoef2005,Yazdchi2011}. Further modifications of the C-K model have been suggested in the form of changes to the Kozeny factor \cite{Astrom1992}, and the replacing particle diamter with the hydraulic diameter \cite{Yazdchi2011,Ozgumus2014}. We observe that FSS permeability does not follow the C-K model, but can be described by a new scaling.

\section{Experiment design}

\subsection{Microfluidic devices}

Each microfluidic channel (see Fig.~\ref{fig:channel}) is fabricated using photolithography. We etch the channel and barrier into a fused silica wafer using reactive ion etching. At each end of the channel, we create access holes by sandblasting with aluminum oxide particles. The upper surface of the channel is sealed with a $\sim15\,\mathrm{\mu m}$ thick layer of cross-linked polydimethylsiloxane (PDMS), which is bonded to a coverslip, and creates the ceiling of the channel. The PDMS and coverslip are removable for repeated reuse of the same channel for multiple experiments \cite{Inglis2004, Inglis2010}. To perform experiments with the microfluidic device, we connect the access holes to milled channels in an acrylic block with o-rings. The acrylic block contains two $20\,\mathrm{\mu L}$ reservoirs for holding suspensions. An aluminum face-plate fastens via screw connectors to complete the seal between the acrylic block and the device and contains a cutaway region to allow visual access to the channel.

As shown schematically in Fig.~\ref{fig:channel}, the height $H$ of the channel is larger than the height $h$ of the barrier. The gap between the top of the channel and the top of the barrier is designed to allow only fluid to flow over and not the particles. We use three different devices for our experiments, each chosen to be slightly larger than the particle diameter. The three devices \cite{Ortiz2013} have channel heights $H = 2.1\,\mathrm{\mu m}$, $1.5\,\mathrm{\mu m}$, $0.94\,\mathrm{\mu m}$, barrier heights $h = 1.7\,\mathrm{\mu m}$, $1.1\,\mathrm{\mu m}$, $0.80\,\mathrm{\mu m}$, and barrier widths $w = 250\,\mathrm{\mu m}$, $200\,\mathrm{\mu m}$, $512\,\mathrm{\mu m}$, and were used for particle diameters $d = 1.04\,\mathrm{\mu m}$, $0.80\,\mathrm{\mu m}$, $0.52\,\mathrm{\mu m}$, respectively. The liquid flux over the barrier is neither parallel nor perpendicular to the barrier; therefore, its net force has both normal (compressive) and tangential (shear) components. The channel width $W$ is at least 4 times larger than the barrier width $w$; therefore, hydrodynamic effects from the side walls are negligible at the barrier.

\begin{figure}
  \includegraphics[width=0.8\linewidth]{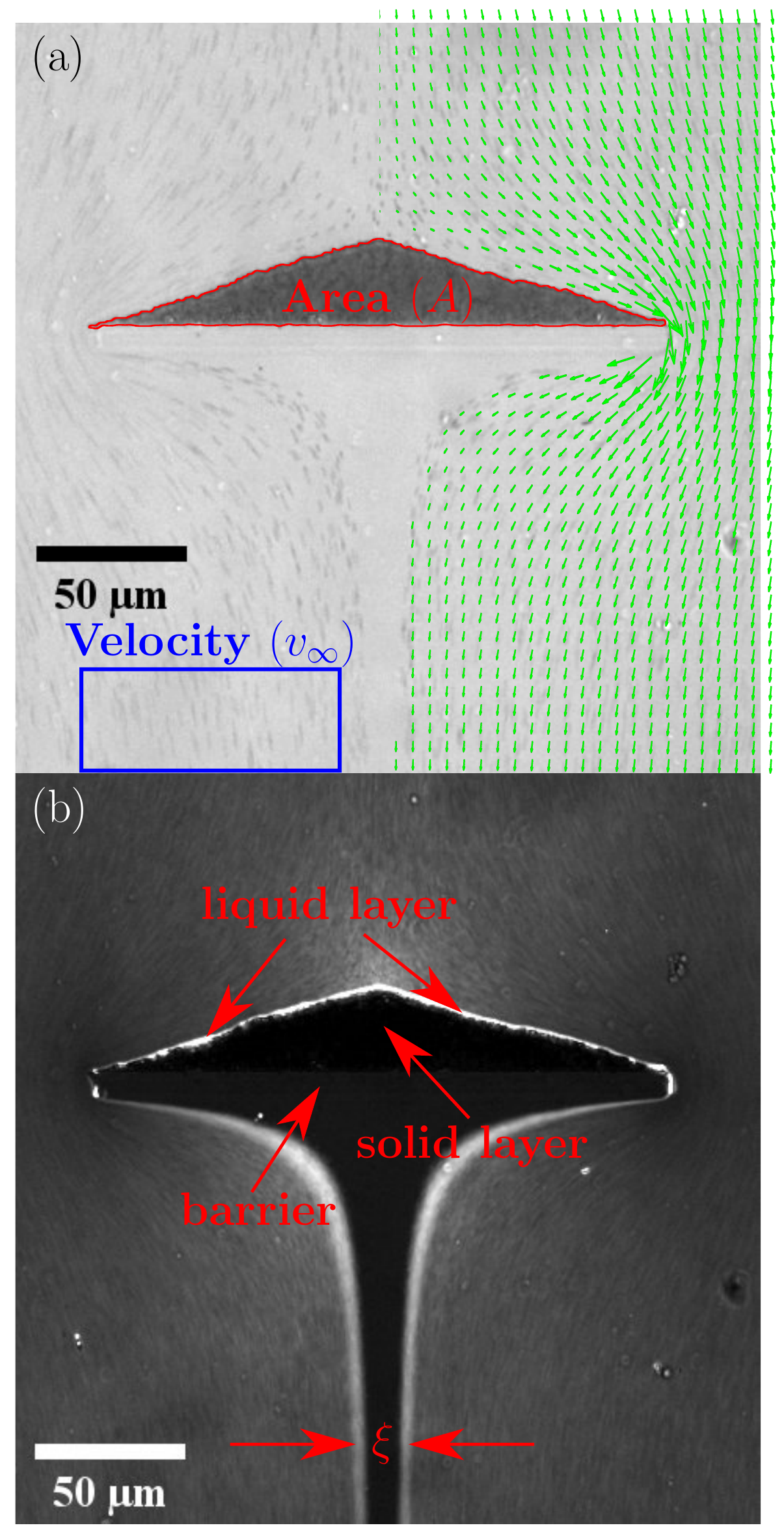}
 \caption[Sample pile formation]{(a) Sample brightfield microscopy image of a steady-state pile, with its boundary outlined in red to define the pile area $A$. The blue rectangle represents the region of interest for PIV calculations, selected to be far from the hydrodynamic influence of the barrier. The green vector field on the right side of the image shows the results of hydrodynamic simulations. (b) Processed image of a steady-state pile, in which each pixel's brightness is proportional to the standard deviation (over time). Bright regions have large intensity variability and dark regions have small intensity variability. The width $\xi$ of the exclusion zone is labeled below the pile.}
 \label{fig:ExMeas}
\end{figure}

The suspension flow velocity is set by pressurized air applied to a single reservoir, which we refer to as the inlet reservoir, via a digital pressure regulator (AirCom PRE1-UA1) and a custom flow control system, operating at 0 to 10 kPa. Flow velocities range from $0.1-100\,\mathrm{\mu m}\cdot \mathrm{s}^{-1}$, with confidence in the stability of low velocities derived from the 10 $\mathrm{ms}$ response time and the high sensitivity of the pressure regulator ($0.1\%$ above the set point). At the highest velocity, the Reynolds number is $10^{-4}$; this indicates that inertial effects are negligible for our experiments.

\subsection{Colloidal stabilization}

The suspension comprises weakly polydisperse ($\sim5\%$) polystyrene microspheres (Bangs Laboratories) of diameters $d = 0.52\,\mathrm{\mu m}$, $0.80\,\mathrm{\mu m}$, $1.04\,\mathrm{\mu m}$, suspended in a citric acid buffer. The polydispersity is sufficient to suppress crystallization within the piles. The buffer's ionic strength is set at $10\,\mathrm{mM}$ in order to remain within the DLVO regime 
 \cite{Bostrom2001}. The pH of the buffer is 6.0, above the isoelectric point (IEP) of all surfaces in the experiment: polystyrene (IEP = 3.4) \cite{Oshawa1986}; silane coating (IEP $\sim 3.7$) \cite{Kolska2013,Preovcanin2012}; 
polydimethylsiloxane (IEP = 3.0) \cite{Kosmulski2001, Kato1987,Schrott2009,Spehar2003,Ren2001}; and silica (IEP = 3) \cite{Kirby2004}. The buffer solution also contains $0.1\%$ ethylenediaminetetraacetic acid (EDTA) as an antimicrobial agent. Finally, we add Triton-X surfactant at 0.1\% (v/v) to the buffer to form a polymeric brush layer on all microparticles and channel surfaces. The brushes suppress colloidal aggregation of the particles, due to the fact that the polymer brushes are longer than the Debye length. The concentration is set so that the microspheres neither aggregate, nor display attraction due to depletion forces.

To limit particle-sticking at the channel walls, we coat them with trichlorosilane (Gelest), a strongly hydrophobic molecule; its contact angle with water is $110^\circ$. To create an even coating, we use atmospheric pressure chemical vapor deposition. Because silane bonds with hydroxyl groups, we first apply a low kinetic energy oxygen plasma to our silica surface to create hydroxyl groups. Next, we flow in nitrogen gas containing (tridecafluoro-1,1,2,2-tetrahydrooctyl) trichlorosilane for 90 minutes, followed by 30 minutes of pure nitrogen for cleaning. Finally, we create a homogeneous silane layer by flowing a lower molecular weight silane ((3,3,3-trifluoropropyl)-trichlorosilane) for 90 minutes, which fills any gaps in the silane layer left by the initial deposition. This creates an anti-stiction layer on all silica surfaces, and provides absorption sites for the Triton-X polymer brushes described above. Trichlorosilane molecules at the top of the silane layer contain a single reactive hydrogen bond, which decreases the silane's anti-stiction property. We combat this reactivity by aging the device in atmospheric conditions for at least 12 hours.

Because the silane layer is sufficiently robust (it is possible to swab away leftover particles at the end of an experiment), we are able to utilize the same silanized device for multiple experiments before reseting the device. When a more thorough cleaning is necesssary, we remove all non-silica materials from the microchannel by submerging the device in Nano-Strip (Cyantek) at $120^\circ$C for two hours. To ensure no surface modifications remain, we briefly re-etch the device using the same protocol used to create it. Because no photoresist is present on the device during the re-etch, all silica surfaces are etching equally, and the microchannel dimensions are maintained. 

\subsection{Imaging}

Because the silica channel is transparent, we make our observations using transmission brightfield microscopy. Therefore, bright regions of images such as Fig.~\ref{fig:ExMeas}a correspond to particle-free zones; such regions appear both in the main channel and in the region below the barrier. Particles in the suspension and in the pile appear dark, although not homogeneously due to being located at various depths with respect to the imaging plane. Note that particles may appear either as dark circles (at low velocities) or as dark streaks (at high velocities). We collect video images at a resolution $512 \times 512$ at $10$~Hz, using an Andor iXon 897 camera.

The image presented in Fig.~\ref{fig:ExMeas}a illustrates an example of an FSS after reaching steady-state. During an equilibration time leading up to this state, piles form on the barrier in a dense, approximately-triangular shape that grows in size as particles are deposited at its surface at a faster rate than they dissipate off. Within individual images, we characterize the geometry of the pile and the flow around it by four quantities: its area $A$, angle of repose $\theta$, the flow velocity $v_\infty$, and the width $\xi$ of the excluded zone. In addition, we use the full series of images to monitor brightness fluctuations within the liquid layer, to characterize its dynamics. 

To determine the pile area $A$, we operate on images to which a 2 pixel wide Gaussian blur has been applied, to lessen the effects of noise. For each column of the image, we find locations of steepest rising and falling slope along the column to determine the location of the edges of the profile. This defines the profile ${\cal H}(x,t)$ defined as a function of column position $x$ and time $t$, and the position of the barrier at $x=0$. We compute the pile area by integrating the height profile across the barrier width; an example outline of ${\cal H}(x)$ is shown in Fig.~\ref{fig:ExMeas}a.

Note that the profile is only approximately triangular: its shape is slightly concave, and has local fluctuations. This deviation from a constant angle of repose is accounted for by calculating its value geometrically, via $\theta = \tan^{-1}\left(\frac{4A}{w^2}\right)$. Note also that the profile's concavity is coincident with the changing orientation of the flow field at its surface, as illustrated in Fig.~\ref{fig:ExMeas}a. 
Thus, the angle of repose differs in its cause from that of an ordinary granular pile, where it arises from the components of gravitational and frictional forces. In these frictionless FSS, $\theta$ characterizes the pile's stability with respect to the interplay of normal and tangential fluid forces. 

To determine the average flow velocity $v_\infty$, we perform particle image velocimetry (PIV) at a location far away from the hydrodynamic effects of the barrier. This region is marked by the rectangle in Fig.~\ref{fig:ExMeas}a. Reported values are averaged over 1 s (10 frames).

Another feature of the flow is the particle-fee exclusion zone which arises in the wake of the barrier. As discussed in \citet{Ortiz2014}, its width $\xi$ (see Fig.~\ref{fig:ExMeas}b) is useful for characterizing the pile's permeability, which we will measure in \S\ref{ssec:Correlations}. 

Last, we characterize the fluctuations of the FSS within the pile by considering the standard deviation of the brightness timeseries, recorded at each pixel within the pile. A sample of such an analysis is shown in Fig.~\ref{fig:ExMeas}b. Here, the dark regions indicate the presence of only small fluctuations in image brightness (the pile, the barrier, and the exlusion zone); the bright regions correspond to regions where particles are changing their positions. In this image, both the FSS (dark triangle) and the liquid-like layer that surrounds it (bright pixels on the upstream side of the pile) are visible. From such images, we determine whether or not a FSS has formed, based on the criteria of finding a measurable region of low fluctuations located upstream of the barrier.

\subsection{Experimental Protocol}

To begin each experiment, we wet the device with a mixture of 50$\%$ buffer solution and 50$\%$ semiconductor-grade methanol, without microspheres. This is necessary due to the strong hydrophobicity of the silane layer; wetting the device with pure buffer solution would require a pressure large enough to break the PDMS seal. We sequentially refill the reservoir with buffer/methanol mixtures of decreasing concentrations of methanol until only buffer solution is present. To avoid having the PDMS-coated coverslip adhere to the barrier due capillary forces from the approaching fluid, we apply a pressure such that we see a slow wetting wavefront ($v_f < 25\,\mathrm{\mu m}\cdot\mathrm{s}^{-1}$) near the barrier. The slow wetting front allows for visual confirmation of whether the PDMS contacts with the barrier.

\begin{figure}
\centering
\includegraphics[width=\linewidth]{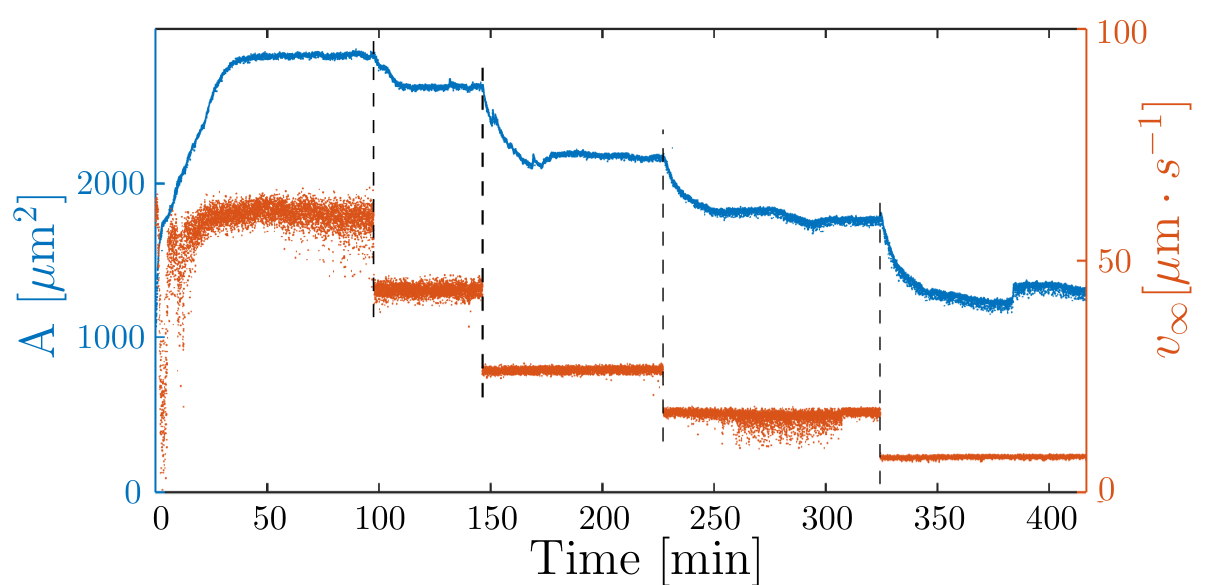}
\caption[Timeseries of FSS area]{Representative time-series of the area $A$ occupied by the FSS (blue, upper curve), and the particle velocity (averaged over 1 s intervals) $v_\infty$ (red, lower curve), for a single experimental run using $800$~nm particles. Each dashed line indicates when a new (decreased) pressure setting was applied to the channel. The fluctuations in $v_\infty$ are an artefact of particle-concentration fluctuations that complicate the use of PIV; the standard deviation of these values is used to set the error bars in Figs.~\ref{fig:FigPecletAngle} and \ref{fig:FigPecletTimes}.} 
\label{fig:ExPile}
\end{figure}

Once the device is wet with buffer solution, we inject the microsphere suspension into the reservoir at a concentration of 0.1\% (w/w). We then apply a constant pressure sufficient that particles flow above a critical velocity. After 1-2 hours, the particles form a steady-state pile at the barrier. From this point, we continue to decrease the applied pressure in steps of $1$~kPa (and therefore also $v_\infty$) and at each step wait for a steady-state pile to form. This sequence is repeated until no pile is formed. This results in values of $v_\infty$ between $0.1\,\mathrm{\mu m}\cdot\mathrm{s}^{-1}$ and $100\,\mathrm{\mu m}\cdot\mathrm{s}^{-1}$, spanning regimes at which the advective and diffusive timescales are of similar magnitude up to advection-dominated flows, as will be discussed in \S\ref{ssec:timescales}.

This protocol is presented in Fig.~\ref{fig:ExPile}, in which the FSS initially grows into a pile of approximately triangular shape and area $A$, until it reaches a steady state (see Fig.~\ref{fig:ExMeas}a). As the pressure is decreased in steps (at each vertical dashed line), the area decreases as well. As the system equilibrates to each lower velocity, particle diffusion off the surface of the pile (erosion) dominates over deposition. Each steady-state pile is characterized by small variations ($\lesssim 5\%$) from the mean area; these are the flat regions of the $A(t)$ graph in Fig.~\ref{fig:ExPile}. We found that the time to reach steady state was approximately invariant with respect to particle size. Note that as the experiment begins, the particle concentration is still in the process of stabilizing and increased fluctuations are present.

\subsection{Timescales \label{ssec:timescales}}

Two timescales are relevant to the dynamics: the advective timescale $\tau_A = \frac{(d/2)}{v_\infty}$ and the diffusive timescale $\tau_D = \frac{(d/2)^2}{D}$, where $D=\frac{kT}{6\pi\eta(d/2)}$ is the diffusion coefficient calculated from the the Stokes-Einstein-Sutherland equation, where $\eta$ is the dynamic viscosity, $k$ is the Boltzmann constant, and $T$ is the temperature. Our choice of particle sizes and flow velocities allows us to access $0.01\,\mathrm{s} <\tau_A < 1\,\mathrm{s}$ and $ 0.1\,\mathrm{s} < \tau_D < 0.5\,\mathrm{s}$. Experimental values of the diffusion differ from our theoretical calculation due to the effects of confinement. These two timescales also define the P\'eclet number of the system, with $\mathit{Pe} = \tau_D / \tau_A$. These calculations neglect two effects: the diffusion constant $D$ is modified by the aspect ratio of the channel, and the flow velocity decreases in and near the FSS.

\section{Results}

\subsection{Angle of repose}

\begin{figure}
\centering
\includegraphics[width=0.9\linewidth]{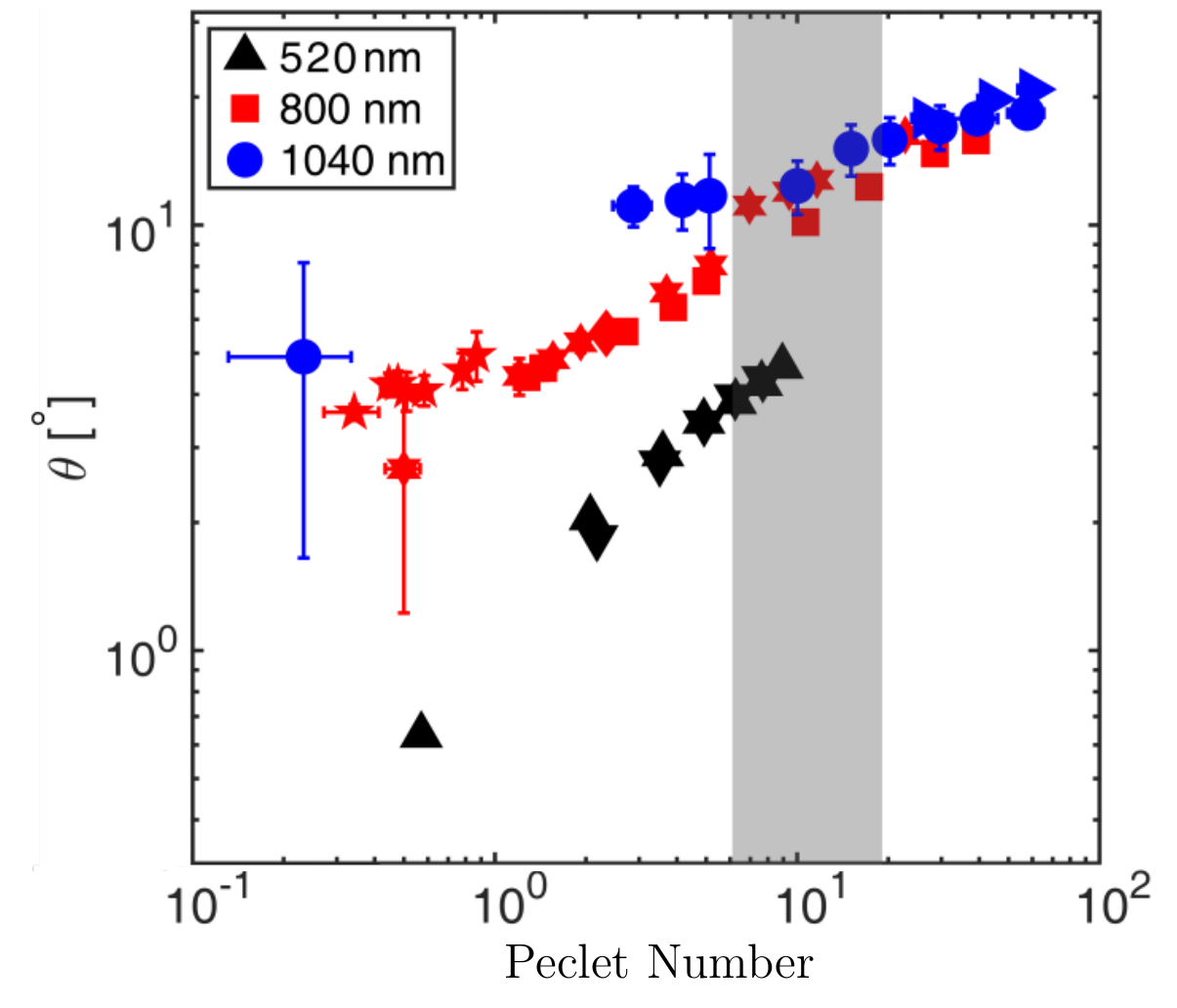}
\caption[P\'eclet Number Analysis]{The pile angle, $\theta$, versus the P\'eclet number. Horizontal error bars represent uncertainty in $v_\infty$ and vertical error bars represent uncertainty in $\theta$. Data for $d=520\,\mathrm{nm}$ are from \citet{Ortiz2013}.}
\label{fig:FigPecletAngle}
\end{figure}

Previously, \citet{Ortiz2013} reported that the angle of repose $\theta$ increased with the P\'eclet number, above a critical value $\mathit{Pe}_c$ required to form a FSS, and that the increase was consistent with an empirical relationship $\theta \propto (\mathit{Pe} - \mathit{Pe}_c)^{1/2}$. First we examine the generality of this finding as a function of the particle diameter $d$. In Fig.~\ref {fig:FigPecletAngle}, we observe that although these findings are qualitatively consistent, neither $\mathit{Pe}_c$, nor the power law is universal. Instead, $\mathit{Pe}_c$ decreases with particle size, meaning that FSS form  more readily at lower flow velocities.

We also observe particle-size dependence in the increase of $\theta$ as a function of $\mathit{Pe}$. First, the angle of repose is systematically larger for larger particles. Second, we observe a distinction between thermal ($\mathit{Pe} \lesssim 10$) and athermal ($\mathit{Pe} \gtrsim 10$) FSS, as marked by the gray bar in Fig.~\ref {fig:FigPecletAngle}. At low $\mathit{Pe}$, there is stronger particle-side dependence than at high $\mathit{Pe}$. Even though there is a power-law relationship, the exponent is not universal and its value decreases with particle size.
For large $\mathit{Pe}$ (the athermal regime), very similar angles of repose are observed, independent of particle size. No data for $d = 520\,\mathrm{nm}$ is in the athermal regime since it would have required applying a higher pressure than the channel could support.

\subsection{Dynamics of the liquid layer \label{ssec:Correlations}}

The angle of repose is primarily set by particles undergoing dynamics within the liquid-like layer near the boundary (see bright regions in Fig.~\ref{fig:ExMeas}b). Pixel-level intensity fluctuations can occur for one of three reasons (1) particle diffusion within the liquid layer, on timescale $\tau_D$; (2) an advective process, characterized by $\tau_A$, which could be dominated either by  migration within the liquid layer due to shear stresses, or particle deposition from the flow due to a flow component normal to the liquid layer; (3) occasional avalanche events that occur quickly, but as discrete events also reported in \cite{Ortiz2014}. In our data analysis, we can identify intervals during which no large-scale avalanches occurred, and focus on the diffusive and advective timescales in order to further characterize the thermal (low $\mathit{Pe}$) and athermal (high $\mathit{Pe}$) regimes.

\begin{figure}
\centering
\includegraphics[width=\linewidth]{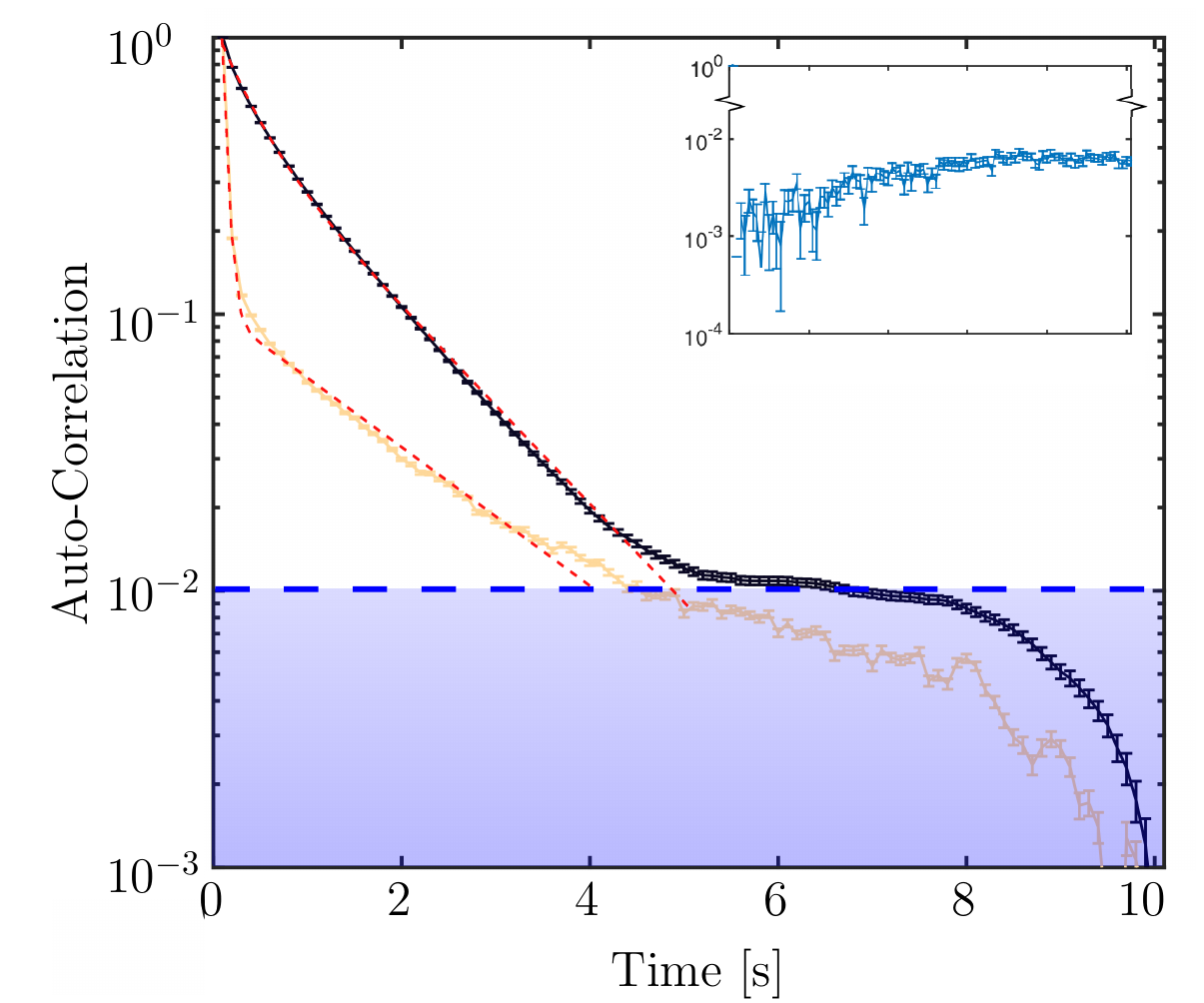}
\caption{Example correlations of the liquid layer of FSS with $800$~nm particles at high flow velocity (yellow), $v_\infty = 59.67\,\mathrm{\mu m\cdot s^{-1}}$, and low flow velocity (black), $v_\infty = 0.52\,\mathrm{\mu m\cdot s^{-1}}$. The dashed blue line is representative of the signal strength of the autocorrelation of the camera background. Data below the dashed blue line (blue region) is not included in our correlation analysis. Example fits following Eq. \ref{eq:DEFit} are shown as dashed red lines. Inset: Autocorrelation of the exclusion zone as a measurement of camera background.}
\label{fig:FigCorr}
\end{figure}

For each pixel within the liquid layer, we record a timeseries of its intensity fluctuations and calculate the temporal autocorrelation function. Sample autocorrelations, averaged over the whole liquid layer, are shown in Fig.~\ref{fig:FigCorr}, for both the athermal (orange) and thermal (black) regimes. In performing the analysis that follows, we discard all data below an autocorrelation value of $10^{-2}$, which is the background correlation level (shown in Fig.~\ref{fig:FigCorr}'s inset).

Motivated by the predicted presence of two timescales (diffusion and advection), we fit each autocorrelation function to a sum of exponentials:
\begin{equation}
\label{eq:DEFit}
 C(t) = c_S\exp{\left(-\frac{t}{\tau_S}\right)} + c_L\exp{\left(-\frac{t}{\tau_L}\right)},
\end{equation}
where $\tau_L$ and $\tau_S$ are assigned to the longer and shorter timescale, respectively. The exponential decays of each timescale are weighted by the magnitudes of $c_L$ and $c_S$, respectively. Sample fits are shown in Fig.~\ref{fig:FigCorr}, in which two distinct decays are apparent for the athermal (high $\mathit{Pe}$) dataset, indicating a separation of timescales. For the thermal (low $\mathit{Pe}$) dataset, the two timescales are indistinct from each other, as expected from the definition of the P\'eclet number.

\begin{figure}
\centering
\includegraphics[width=0.9\linewidth]{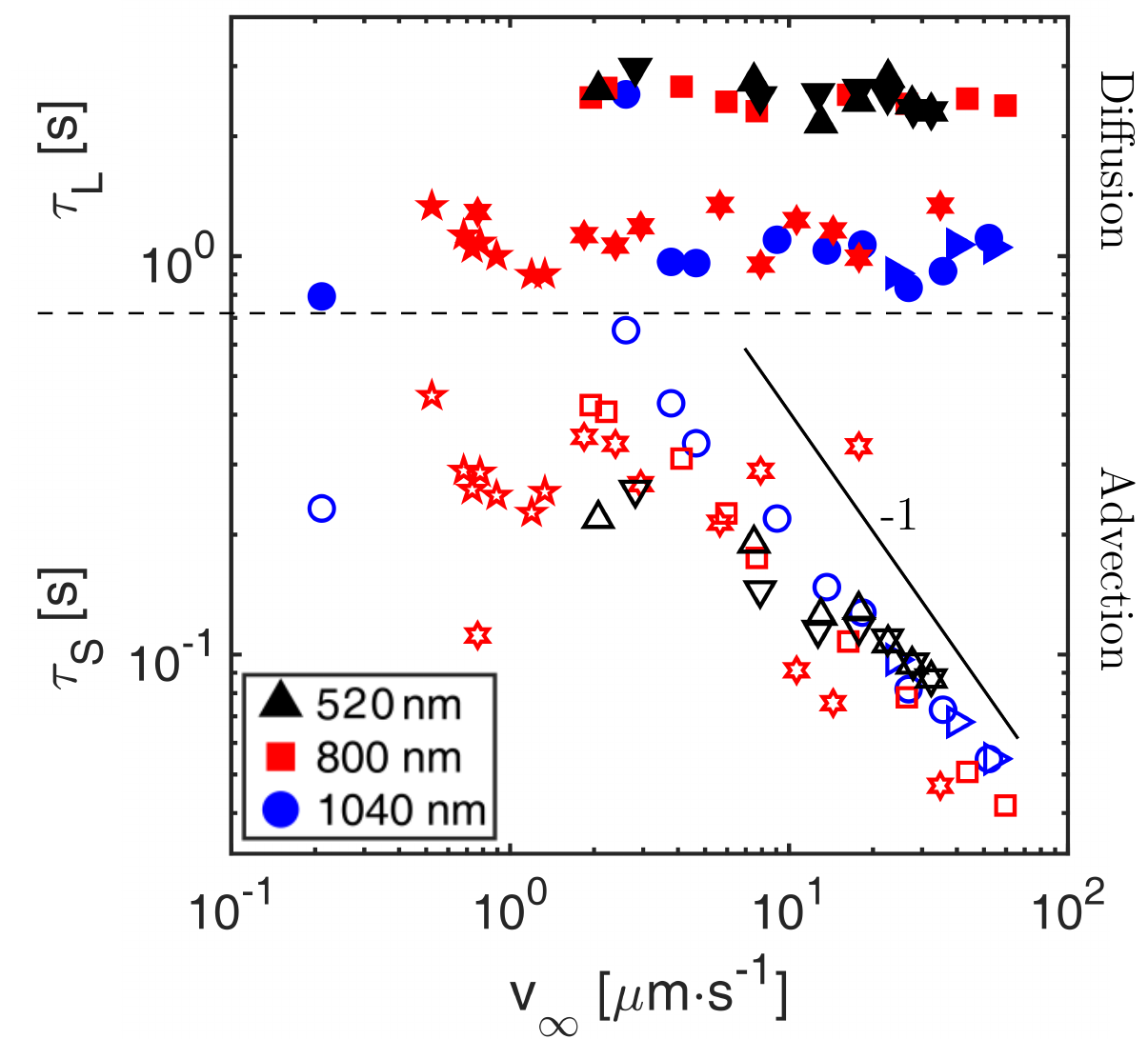}
\caption[Autocorrelation]{Values of $\tau_L$ (solid symbols) and $\tau_S$ (open symbols) obtained by fitting the temporal autocorrelation functions to Eq.~\ref{eq:DEFit}. We associate $\tau_L$ with the diffusion timescale, and $\tau_S$ with the advective timescale.}
\label{fig:timescales}
\end{figure}

To associate each of the two $(\tau_L, \tau_S)$ timescales with the physical process, we plot their values as a function of $v_\infty$ (see Fig.~\ref{fig:timescales}). We observe that $\tau_L$ is independent of $v_\infty$ and depends only on particle size; therefore, we associate $\tau_L$ with $\tau_D$. The difference in $\tau_L$ values for the 800\,nm experiments is caused by different particle concentrations between experiments. At higher particle concentration, the diffusive time scale is shorter. Additionally, we observe that at higher particle concentration in the free flow, the liquid layer has an increased thickness. We thus speculate that the diffusive time scale $\tau_D$ is a function of the free particle concentration since a thicker liquid layer creates higher local pressure and density, which should decrease the effective mobility of individual particles.

\begin{figure}
\centering
\includegraphics[width=0.9\linewidth]{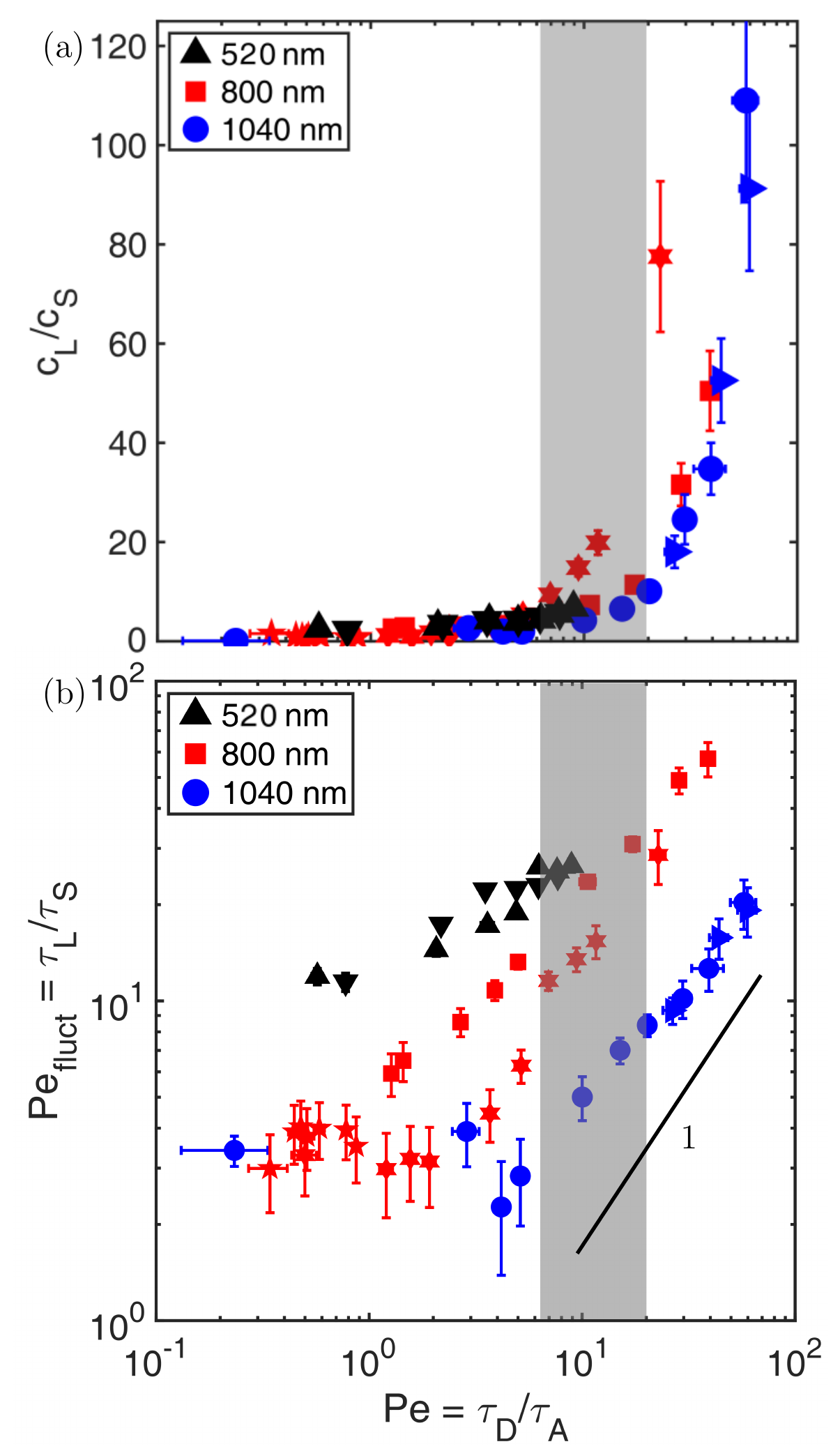}
\caption[P\'eclet Number Analysis]{(a) P\'eclet number versus the amplitudes associated with the short and long time scale exponentials in Eq. \ref{eq:DEFit}. Horizontal error bars represent uncertainty in the measurement of $v_\infty$ and vertical error bars represent uncertainty in fluctuation correlations and in our fit model. (b) Ratio of short and long time scales versus the P\'eclet number. The gray bar in each plot illustrates the transitional regime from thermal to athermal behavior. Data for $d=520\,\mathrm{nm}$ are from \citet{Ortiz2013}.}
\label{fig:FigPecletTimes}
\end{figure}

In addition, $\tau_S$ decays as $v_\infty^{-1}$ as expected, and therefore can be associated with $\tau_A$. There is no apparent particle-size dependence in $\tau_S$. Note that the two timescales do not contribute equally to the the decay of correlations. As shown in Fig.~\ref {fig:FigPecletTimes}a, the amplitude ratio $c_L/c_S$ is $\approx 1$ in the thermal regime, where both advection and diffusion contribute to the particle-scale dynamics within the liquid layer. In the athermal regime, the longer timescale (now associated with $\tau_D$) dominates over diffusion, as expected. This transition near $\mathit{Pe} \approx 10$ corresponds to the transition to universal angles of repose shown in Fig.~\ref{fig:FigPecletAngle}. 

These observations suggest that we define a P\'eclet-like number from the ratio $\mathit{Pe}_\mathrm{fluct} = \tau_L/\tau_S$, in which the numerator is again associated with diffusion, and the denominator with advection. As shown in Fig.~\ref {fig:FigPecletTimes}b, the particle-scale $\mathit{Pe}_\mathrm{fluct}$ is approximately proportional to the one calculated from the known properties of the suspension and flow, but only for sufficiently high $\mathit{Pe}$. At low $\mathit{Pe}$ we observe little variation in the values of $\mathit{Pe}_\mathrm{fluct}$. We further observe that there is an unexplained particle-size dependence in the relationship between $\mathit{Pe}_\mathrm{fluct}$ and $\mathit{Pe}$: for smaller particles, this dependence has a lower exponent than for larger particles.


\subsection{Pile permeability \label{ssec:PilePerm}}

From observations of the changing width $\xi$ of the exclusion zone (see Fig.~\ref{fig:ExMeas}b), we can infer that the hydrodynamic permeability of the FSS ($\kappa_\text{FSS}$) changes in response to both the particle size and the flow velocity. To measure the permeability, we utilize the approach of  \citet{Ortiz2014}, in which $\kappa_\text{FSS}$ is assumed to be homogeneous and the FSS has a isosceles triangular shape. Using the MATLAB PDETool solver, we are able to identify the value of $\kappa_\text{FSS}$ at which the simulated $\xi$ matches the experimentally-observed value.
In the numerical simulations, the permeability $\kappa\left(x,y\right)$ is the coefficient of the Laplace equation $\nabla\cdot[\kappa(x,y)\nabla p] = 0$; solutions provide the fluid's flux field $\vec j\left(x,y\right) =\kappa\nabla p$, which can be directly compared to video at all locations except for inside the FSS and over the barrier. In writing $\kappa(x,y)$, we include contributions from $\kappa_\text{FSS}$ as well as the channel ($\kappa_\text{channel}$) and the barrier ($\kappa_\text{barrier}$).

\begin{figure}
\centering
\includegraphics[width=\linewidth]{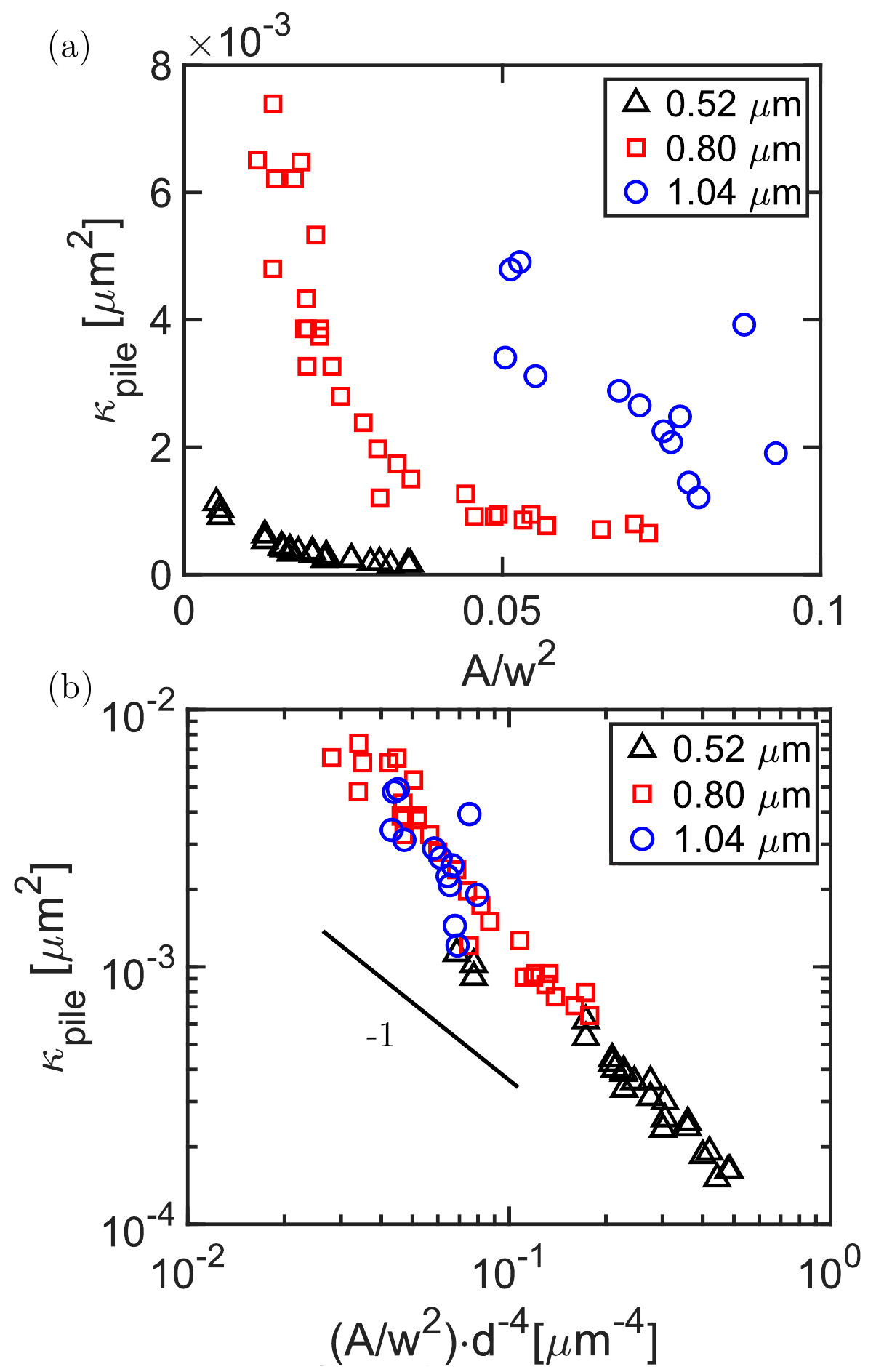}
\caption[Permeability Analysis]{(a) Pile permeability $\kappa_{\mathrm{FSS}}$ versus normalized pile area $A/w^2$. The 3 particle sizes each display similar behavior of exponential decrease in $\kappa_{\mathrm{FSS}}$ as $A/w^2$ increases. (b) Pile permeability as a function of scaled, normalized pile area. A reference slope of $-1$ is shown below the data points. Data for $d=520\,\mathrm{nm}$ are from \citet{Ortiz2013}.}
\label{fig:PilePerm}
\end{figure}

A sample output for the velocity field ${\vec v}(x,y)$ is shown as an overlay in Fig.~\ref {fig:ExMeas}a, associated with the value of $\kappa_\text{FSS}$ for which $\xi$ was the best fit.
We repeat this calculation for the steady-state pile at each pair of $(d,v_\infty)$ values. 
As shown in Fig.~\ref{fig:PilePerm}a, the value of $\kappa_{FSS}$ varies over about an order of magnitude for each particle size, and over two orders of magnitude for the entire dataset. Ortiz et al. \cite{Ortiz2014,Ortiz2016} reported that particles in an FSS are in intermittent contact and thus the packing density is dependent on the stress. Since the area is a monotonic function of the stress exerted by the flow, we thus should anticipate the shape of curves in Fig.~\ref{fig:PilePerm}a on the basis of the Carman-Kozeny argument for flow through porous media \cite{Carman1937,Kozeny1927}. 

Empirically, we find that we can collapse the data onto a single power-law type curve by rescaling the abscissa in Fig.~\ref{fig:PilePerm}a ($A/w^2$) by the dimension $d^{-4}$ (Fig.~\ref{fig:PilePerm}b). This is a remarkable statement since we did not find a universal relationship between $\mathit{Pe}$ and $A/w^2$, and should not expect a universal relationship between $\mathit{Pe}$ and the packing fraction either. We discuss some possible explanations below.


\section{Discussion}
We have expanded the understanding of FSS at equilibrium, demonstrating that the particle size has a complex influence on the conditions under which FSS form, the size of a FSS as function of the P\'eclet number, and fluctuations of the liquid-like layer at the interface between the FSS and the flow. The dominant feature is a transition between diffusion-dominated and shear-dominated dynamics at that interface, which we named the thermal to athermal transition. In completely athermal (granular) systems under cyclic shear, compaction and even crystallization are observed \cite{Pouliquen2003,Rietz2018}. While prior work \cite{Ortiz2013} used a critical P\'eclet number $\mathit{Pe_c}$ to characterize the onset of pile formation, we have not found a universal $\mathit{Pe_c}$ for particle sizes. 

The thermal to athermal behavior is seen most clearly in the time-autocorrelations of fluctuations within the liquid layer, which undergo distinct changes as the P\'eclet number is varied. As introduced in Eq. \ref{eq:DEFit}, fluctuation autocorrelations can be satisfactorily described by a sum of two exponential decays, which introduces two main time scales by which our system is governed. Specifically, the liquid layer has time scales associated with thermal fluctuations, bulk flow due to shear, and deposition and erosion events.  The velocity dependence of the shorter time scale identifies it either with shearing within the boundary or deposition/erosion events. The lack of dependence of this time scale on the particle density impacting the FSS indicates that the time scale is connected to a shearing process. The longer time scale is attributable to diffusion since it is flow-independent. We find that the P\'eclet number provides a good estimator of the transition since it is the ratio of time scales of the thermal fluctuations and advection. 

We have observed an interesting similarity in the observed angle of respose once piles have entered the athermal regime for the two larger particles; this behavior is distinct from the thermal behavior of the smaller particles. We speculate that a similar transition could occur in microfiltration applications. Our semi-permeable barrier acts as a membrane for the fluid flowing in our system stopping the ``foulant'' colloids. Our system is particularly similar to setups in cross-flow filtration where the liquid flow has both normal and tangential components with respect to the membrane. As membranes begin to foul, they form cakes, which are build-ups of solid-like particulate and colloidal matter. In this framework, FSS are a type of filter cake that could either fall into a thermal regime ($\mathit{Pe}<1$) where diffusion competes with advection and a thermal-like solid is formed, or an athermal regime ($\mathit{Pe}>10$) where a colloidal solid is formed that is dominated by macroscopic quantities like shear stresses similar to a granular solid.

Critical fouling has been defined with respect both the initial onset of solid (cake) formation (critical flux) \cite{Field1995,Howell1995,Wu1999,Field2011}, as well as the point at which the cake is irreversible (i.e. if the flow is turned off, the solid remains) \cite{Aimar2008}. Since all FSS formation reported in this manuscript is reversible, we propose that the parameter space between these two limits should be divided into two distinguishable regimes.

Note that filtration is typically characterized in terms of flow through the filter cake, and not the simple the presence of the cake. The measurements provided here show that the permeability of the FSS or filter cake just beyond the critical filter flux is a strong function of the $\mathit{Pe}$, and thus the flux itself. In particular, the permeability drops rapidly with increasing $\mathit{Pe}$, and the particle size dependence is strong.

However, we have also noted that the Carman-Kozeny model cannot account for the entire functional relationship. Importantly, the reduced area or pile angle relationship cannot be motivated using the model. Furthermore, the scaling of the permeability $\kappa$ with $d^{-4}$ is in clear contradiction to our expectation of a scaling of $d^{-3}$. We clearly can identify three weaknesses of the model.
First, our device geometry, with close to one monolayer of colloidal particles, does not match the common assumption of a porous solid that is semi-infinite. Such effects have been observed in recent work on microstructure and stress \cite{Lin2016}, colloidal assembly \cite{Jiang2014}, and active particle diffusivity \cite{Wang2014}. Note that there is a possible dependence of the permeability on the channel height, as the collapse in Fig.~\ref{fig:PilePerm}b can also be achieved if the abscissa is changed from $(A/w^2)\cdot d^{-4}$ to $(A/w^2)\cdot d^{-3} H^{-1}$.
The second shortcoming of our model is that it assumes that the FSS has a homogeneous porosity, which is not strictly correct. In particular the stress on particles increases on a path from the free flow interface to the barrier interface, and we thus also expect a gradient in porosity. The last shortcoming is the neglect of electroviscosity, which could arise at the considered length scales of pores formed by the colloids.

\paragraph*{Acknowledgements} We are grateful to the Triangle MRSEC under grant number NSF DMR-11-21107, for funding this work. The fabrication of the devices was done at NC State's Nanofabrication Facility and Duke's Shared Materials Instrumentation Facility. The development of the silanization technique was developed through helpful conversations with Laura Clarke.

\bibliography{mainbib}

\end{document}